\documentclass[a4paper,pre,amsmath,amssymb,preprint]{revtex4}

\usepackage{graphicx}

\DeclareMathOperator{\sech}{sech}
\DeclareMathOperator{\Real}{Re}
\DeclareMathOperator{\Imag}{Im}

\newcommand{\conj}[1]{{#1}^{\ast}}
\newcommand{\order}[1]{{\mathcal O}\left(#1\right)}
\newcommand{\be}{\begin{equation}}
\newcommand{\ee}{\end{equation}}

\begin{document}
\title{Asymptotic expansion of the wobbling kink}
\author{O.F. Oxtoby}
\email{Oliver.Oxtoby@gmail.com}
\affiliation{
CSIR Computational Aerodynamics, Building 12, P.O.~Box 395, Pretoria
0001, South Africa
}
\author{I.V. Barashenkov}
\email{igor@odette.mth.uct.ac.za; Igor.Barashenkov@uct.ac.za}
\affiliation{Department of Maths and Applied Maths, University 
of Cape Town, Rondebosch 7701, South Africa}
\date{Last update on \today}

\begin{abstract}
The method of multiple scales is used to study the 
wobbling kink  of the $\phi^4$ equation. 
The amplitude of the wobbling is shown to decay very slowly, as
$t^{-1/2}$, and hence the wobbler turns out to be an
extremely long-lived object. 
\end{abstract}

\maketitle

\section{Introduction}

Since the early 1960s, the one-dimensional $\phi^4$ theory 
has been among the most heavily utilised models 
of statistical mechanics and
condensed-matter physics \cite{Stats}. At the same time,
it served as a testing ground
for a variety of ideas in topological quantum field theory
\cite{QFT_books} and cosmology \cite{cosmology}.
 The equation of motion for the model reads
\begin{align}
\label{phi4}
\frac{1}{2}\phi_{tt}-\frac{1}{2}\phi_{xx} - \phi + \phi^3 = 0,
\end{align}
and the fundamental role in  applications is played by
its kink solution, 
 \begin{align}
 \phi(x,t) = \tanh x.
 \label{tanh}
 \end{align}
Mathematically, the $\phi^4$ kink has a lot in common with its
sine-Gordon counterpart; the two kinks are the 
simplest examples of topological solitons in one dimension.
There is an important difference though; the sine-Gordon equation
is integrable whereas the $\phi^4$ theory is not. 
Accordingly, the kink-antikink interaction becomes a nontrivial 
matter in the $\phi^4$ case \cite{Aubry,collisions,Getmanov}.
Another (not unrelated) difference is that
unlike the kink of the sine-Gordon equation, the $\phi^4$ kink
 has an internal mode --- an extra degree
of freedom which allows for oscillations in the width of the kink.
Although these oscillations are accompanied by the emission of radiation 
(a manifestation of the nonintegrability of the $\phi^4$ model),
the radiation seems to be quite weak and the oscillations are sustained over
long periods of time.  For small oscillation amplitudes, 
this periodically expanding and contracting kink
(termed {\it wobbling\/} kink in literature)
 can be characterised simply as a linear perturbation 
of the stationary kink \eqref{tanh};
however for larger amplitudes one needs a self-consistent 
fully-nonlinear description. 

The wobbling kink was discovered in the early numerical experiments of
Getmanov \cite{Getmanov} who interpreted it as a bound state 
of three ordinary kinks. (For a more recent series
 of numerical simulations,   
see \cite{Belova}.)
 Rice and Mele have reobtained this nonlinear
excitation within
a variational approach employing the width of the kink 
as a dynamical variable \cite{RM}.
Segur then constructed the 
wobbler
as a regular perturbation expansion in powers of the 
 oscillation amplitude \cite{Segur}. He has calculated 
  the first two orders of the perturbation
series and noted the likely occurrence of unbounded terms at 
the third, $\epsilon^3$-, order, implying the consequent breakdown
of the expansion. Subsequently, 
Sukstanskii
and Primak \cite{Sukstanskii} examined the mobility of
the wobbler using
 a variant of the Lindstedt-Poincar\'e technique where
the velocity of the 
kink is adjusted so as to eliminate secular terms at the lowest
orders of the perturbation expansion.  
Although this approach provides valuable insights 
at the lowest orders, it is not suitable to deal with the
 secular terms at the $\epsilon^3$-order 
 and therefore gives rise to a nonuniform expansion.
  Finally,  Kiselev 
studied the perturbed 
$\phi^4$ kink using the Krylov-Bogoliubov 
method \cite{Kiselev}. 
His two papers have a wealth of useful formulas; however a
self-consistent system of equations 
for the kink's parameters was not derived in \cite{Kiselev} 
and the long-term evolution of the 
wobbling kink has remained unexplored.

The aim of this paper is to 
construct the wobbling kink as a singular perturbation expansion 
which remains uniform for long times and large distances.
Our approach employing a sequence of space and time scales will also yield 
equations for the 
 amplitude of the wobbling mode
 which will be used to analyse its long-term evolution.

\section{Multiscale Expansion}
\label{Free}

Making the change of variables $(x,t) \to (\xi, \tau)$, where
$\xi = x - \int_0^t v(t') dt'$,  $\tau =t$,
 we
transform Eq.\eqref{phi4} to the co-moving frame:
\begin{align}
\label{phi4converted}
\frac{1}{2}\phi_{\tau\tau} - v\phi_{\xi\tau} - 
\frac{v_\tau}{2} \phi_\xi - \frac{1-v^2}{2}\phi_{\xi\xi} - \phi +
\phi^3 = 0.
\end{align}
Like the authors
of \cite{Sukstanskii}, we shall determine the kink's
velocity $v(t)$ by imposing  the condition
that the kink be always centred at $\xi=0$  [i.e. at $x=\int_0^t v(t')
dt'$].

At first glance, the inclusion of the function $v(t)$ is unnecessary:
having constructed a quiescent
wobbling kink, we could make it move  simply 
by a Lorentz boost. By introducing $v$ 
explicitly in Eq.\eqref{phi4converted} we wish to check whether 
the wobbling kink can drift with a 
{\it nonconstant\/} velocity. The soliton moving with a variable $v(t)$ 
could obviously not
be Lorentz-transformed to the rest frame. 

We expand the field about the kink
$\phi_0 \equiv \tanh \xi$:
\begin{align}
\label{phiexpans}
\phi &= \phi_0 + \epsilon\phi_1 + \epsilon^2\phi_2 + \ldots.
\end{align}
Here $\epsilon$ is a (formal) small parameter. 
Substituting  \eqref{phiexpans} in \eqref{phi4converted}
and setting to zero coefficients of like powers of $\epsilon$
would constitute Segur's approach which is expected to produce 
secular terms in the expansion \cite{Segur}. To avoid these,
we  introduce  stretched space
and time scales
\begin{equation}
X_n \equiv \epsilon^n \xi, \quad
T_n \equiv \epsilon^n \tau, \quad n=0,1,2,...,
\end{equation}
with the standard short-hand notation for the derivatives:
$\partial_n   \equiv   \frac{\partial}{\partial X_n}$, 
$D_n   \equiv \frac{\partial}{\partial T_n}$.
In the limit $\epsilon \to 0$,
the different scales become uncoupled and may be treated as
independent variables: $\phi_n=\phi_n(X_0,X_1,...; T_0,T_1,...)$. 
We also assume that $v$ is small and slowly varying,  
i.e. $v = \epsilon V$  where
$V = V(T_1, T_2, \ldots)$ is of order 1.
Expanding $\frac{\partial}{\partial \xi}$ and $\frac{\partial}{\partial
\tau}$ using the chain rule, and substituting along with the series
\eqref{phiexpans} in 
Eq.\eqref{phi4converted},
we equate coefficients of like powers of $\epsilon$.

\subsection{Linear corrections}

At $\order{\epsilon^1}$, we obtain the linearisation of Eq.\eqref{phi4}
about the kink $\phi_0=\tanh X_0$:
\be \label{linear}
\frac{1}{2}D_0^2\phi_1 + {\mathcal L}\phi_1 = 0,
\ee
where we have introduced the Schr\"odinger operator
\begin{equation}
{\mathcal L} = -\frac{1}{2}\partial_0^2-1+3\phi_0^2 = -\frac{1}{2}\partial_0^2+2-3\sech^2X_0.
\end{equation}
The general solution of the variable-coefficient
 Klein-Gordon equation 
 \eqref{linear}
can
be written as
\be
\phi_1= C_0 y_0(X_0)
+ A e^{i \omega_0 T_0} y_1(X_0) + c.c. + \phi_R(X_0,T_0),
\label{lin_cor} 
\ee
\be 
\phi_R= 
\int_{-\infty}^{\infty} 
\left[ {\cal R} (p)e^{i \omega(p) T_0}+ {\cal R}^*(-p)e^{-i \omega(p) T_0}
\right]
 y_p(X_0) dp. 
 \label{int_rad}
 \ee
 Here $y_0$ and $y_1$ are eigenfunctions of the operator $\cal L$
associated with eigenvalues $0$ and $\frac32$, respectively:
\begin{eqnarray}
y_0(X_0)= \sech^2X_0; \label{y0} \\
y_1(X_0)= \sech X_0 \tanh X_0. \label{y1} 
\end{eqnarray}
The functions $y_p(X_0)$ are solutions pertaining to
the continuous spectrum of $\cal L$:
\be
{\cal L} y_p(X_0)= \left( 2+ \frac{p^2}{2} \right) y_p(X_0);
\ee
 these were constructed by Segur \cite{Segur}:
\begin{equation}
\label{segurssolutions}
y_p = e^{ipX_0}\Bigg[1+\frac{3(1-ip)}{1+p^2}\tanh X_0 (1+\tanh X_0) 
- \frac{3(2-ip)}{4+p^2}(1+\tanh X_0)^2\Bigg].
\end{equation}
The internal mode frequency $\omega_0= \sqrt{3}$, 
while the phonon frequencies $\omega(p)$ are given by
$\omega(p)=\sqrt{4+p^2}>0$. Finally, 
the coefficients ${\cal R}(p)$ and $A$ are complex; $C_0$ is real,
and $c.c.$ in \eqref{lin_cor}
 stands for the complex conjugate of the immediately
preceding term.

We set the coefficient of the 
translation mode to zero as the kink is assumed to be centered at $X_0=0$. 
Since we are interested in the wobbling of the kink 
sustained over long periods of time, we also set 
the radiation amplitudes ${\cal R}(p)=0$.    
 As a result, the first-order perturbation comprises only the wobbling mode:
\begin{equation}
\phi_1 =
 A(X_1, \ldots; T_1, \ldots)
 \sech X_0\tanh X_0e^{i \omega_0 T_0} + c.c.
 \label{phi_1}
\end{equation}

\subsection{Quadratic corrections} 

At the second order in the perturbation expansion we arrive
at the
equation 
\be
\frac{1}{2}D_0^2\phi_2 + {\mathcal L}\phi_2 
= F_2(X_0,...; T_0,...),
\label{qua}
\ee
where the forcing term is
\begin{subequations} \label{F2}
\begin{equation}
F_2= (\partial_0\partial_1 - D_0 D_1)\phi_1 
- 3\phi_0\phi_1^2+ VD_0\partial_0\phi_1 
+ \frac{1}{2}D_1V\partial_0\phi_0-\frac{1}{2}V^2\partial_0^2\phi_0.
\label{F2a}
\end{equation}
Substituting for $\phi_0$ and $\phi_1$, this  becomes
\begin{eqnarray}
 F_2 = -6|A|^2\sech^2X_0\tanh^3X_0 + \frac{1}{2}D_1V \sech^2X_0
- V^2 \sech^2X_0 \tanh X_0 \nonumber \\
 + \left[ (\partial_1A + i\omega_0 VA) (2\sech^3X_0 
 -\sech X_0) 
- i \omega_0 D_1A\sech X_0\tanh X_0 
 \right]e^{i \omega_0 T_0} + c.c. \nonumber \\
 - 3A^2\sech^2 X_0\tanh^3X_0e^{2i \omega_0 T_0} + c.c. \ \
\label{eps2eqn} 
\end{eqnarray}
\end{subequations}
The $T_0$-independent term in Eq.\eqref{eps2eqn}
and the term proportional to $e^{i\omega_0 T_0}$
are resonant with the two discrete eigenmodes
of the operator in the left-hand side of \eqref{qua}, while
the term proportional to $e^{2i\omega_0 T_0}$
is resonant with its continuous spectrum.  
The latter part of the forcing is localised in the region near 
the origin
and acts as a source of radiation which spreads outward from there.

Once transients have died out, the $\epsilon^2$-correction will consist only of 
the harmonics present in the forcing, i.e.
\begin{align}
\phi_2 = \varphi_2^{(0)} + \varphi_2^{(1)}e^{i\omega_0 T_0} + c.c. 
+ \varphi_2^{(2)}e^{2i\omega_0 T_0} + c.c.,
\label{phi2m}
\end{align}
where $\varphi_2^{(0)}$, $\varphi_2^{(1)}$ and
$\varphi_2^{(2)}$  are functions of 
$X_0$ which satisfy
\begin{equation}
{\mathcal L}\varphi_2^{(0)} = -6|A|^2\sech^2X_0\tanh^3X_0 
+ \frac{1}{2}D_1V\sech^2X_0 - V^2\sech^2X_0\tanh X_0,
 \label{ph20} \ee \be
({\mathcal L}-\tfrac{3}{2})\varphi_2^{(1)} =  \; 
(\partial_1A+ i\omega_0 VA)(2\sech^3X_0-\sech X_0) 
- i \omega_0 D_1A\sech X_0\tanh X_0, 
 \label{ph21} 
 \ee
 and
 \be
\label{linearspreading}
({\mathcal L}-6)\varphi_2^{(2)} = - 3A^2\sech^2X_0\tanh^3X_0.
\ee

These equations
admit bounded solutions if and only if
their right-hand sides are orthogonal to
the corresponding eigenfunctions of the operator
$\cal L$, Eqs.\eqref{y0} and \eqref{y1}.
For this to be the case, we must set $D_1V = 0$ and $D_1A = 0$.
The variation of parameters yields then
\be
\varphi_2^{(0)} = 2 |A|^2 \sech^2X_0 \tanh X_0 
+ \left( \frac{V^2}{2}-3|A|^2 \right)
X_0\sech^2X_0 \label{varphi2}
\ee
and
\be
\label{phi2firstharm}
\varphi_2^{(1)} = -(\partial_1A + i\omega_0 VA)X_0\sech X_0\tanh X_0.
\ee

Although the function $\varphi_2^{(1)}$ decays
to zero as $|X_0| \to \infty$, the product
 $\epsilon \varphi_2^{(1)}$ becomes greater than 
the first-order correction $y_1(X_0)$ for each fixed $\epsilon$
and sufficiently large $|X_0|$.
Consequently, the term $\epsilon^2 \phi_2$ in the expansion \eqref{phiexpans}
becomes greater than the previous term, $\epsilon^1 \phi_1$,
leading to the nonuniformity of the expansion.
In order to obtain a uniform expansion, we 
 set
\be
\partial_1 A + i\omega_0 VA = 0,
\ee
which gives
$A= A_0 e^{-i \omega_0 V X_1}$, with
$A_0=A_0(X_2,X_3,...;T_2,T_3,...)$.

We also note the terms proportional to 
$X_0 \sech^2 X_0$ in Eq.\eqref{varphi2}. These terms
do not grow bigger than the previous term, $\phi_0= \tanh X_0$, yet
they become larger than the difference 
$\phi_0-1$ as $X_0 \to \infty$ and $\phi_0-(-1)$ as $X_0 \to -\infty$. 
If we attempted to construct the asymptotic expansion 
of the function $\phi-1$ at the right infinity 
or the function $\phi+1$ at the left infinity, the terms
in question would cause the nonuniformity of these.
Since the function $X_0 \sech^2 X_0$
is nothing but the derivative of $\tanh (k X_0)$ with respect to $k$,
these terms represent the variation of the kink's
 width. Hence the potential nonuniformity of
 the expansion can be avoided simply by incorporating 
 them in the variable width [see Eq.\eqref{phi_sum} below].
 
We now turn  to  the remaining nonhomogeneous equation,
Eq.\eqref{linearspreading}.
The variation of parameters gives
\be
\label{wrongsoln}
\varphi_2^{(2)} = A^2 f(X_0), \ee
\begin{multline}
f(X_0) = \frac{1}{8} \big[ 6\tanh X_0 \sech^2 X_0 
+ (3-\tanh^2 X_0 +ik_0 \tanh X_0) (\conj{J}_2-J_2^{\infty}) e^{ik_0 X_0} \\
+ (3-\tanh^2 X_0 -ik_0 \tanh X_0)J_2 e^{-ik_0X_0} \big].
\label{f1}
\end{multline}
Here the wavenumber $k_0=\sqrt{8}$;
the function  $J_2=J_2(X_0;k_0)$ is defined by the integral
\be
\label{Jdef}
J_2(X_0;k) = \int_{-\infty}^{X_0} e^{ik\xi}\sech^2 \xi \; d\xi,
\ee
and the constant  $J_2^{\infty}$ is the
asymptotic value of $J_2(X_0;k_0)$ as $X_0 \to \infty$.

The solution \eqref{wrongsoln} describes right-moving radiation for positive 
$X_0$ and left-moving radiation for negative $X_0$.
The function \eqref{wrongsoln} is bounded 
but does not decay to zero as $|X_0| \to \infty$.

\subsection{The amplitude equation}

Collecting terms of order $\epsilon^3$ gives the equation
\begin{subequations}
\begin{equation}
\frac{1}{2}D_0^2\phi_3 + {\mathcal L}\phi_3 =F_3,
\end{equation}
where
\begin{multline}
\label{epscubed}
F_3 = (\partial_0\partial_1 - D_0 D_1)\phi_2 + (\partial_0\partial_2 - D_0 D_2)\phi_1
+ \frac{1}{2}(\partial_1^2-D_1^2)\phi_1 -\phi_1^3 - 6\phi_0\phi_1\phi_2 + VD_0\partial_0\phi_2 \\
+VD_0\partial_1\phi_1 
+ VD_1\partial_0\phi_1 
+ \frac{1}{2}D_2V\partial_0\phi_0 -\frac{1}{2}V^2\partial_0^2\phi_1.
\end{multline}
\end{subequations}
Having evaluated $F_3$ using the known functions $\phi_0$,
$\phi_1$ and $\phi_2$, we again decompose the solution into 
 simple harmonics as we did at $\order{\epsilon^2}$.
The solvability condition for
the zeroth harmonic in equation \eqref{epscubed}
gives $D_2V = 0$, which means that $V$ remains constant
up to times $t \sim \epsilon^{-3}$. 
The solvability condition for 
the first harmonic produces
\begin{align}
\label{freeamp}
 i \frac{2  \omega_0 }{3} D_2A + \zeta |A|^2A - V^2A = 0,
\end{align}
where
\be
\zeta =6
\int_{-\infty}^{\infty} \sech^2X_0\tanh^3X_0\Big[\frac{5}{2}\sech^2X_0\tanh
X_0 
 -3X_0\sech^2X_0 + f(X_0)\Big]\,dX_0. \  \
\ee
The imaginary 
part of $\zeta$ can be evaluated analytically:
\be
\Imag \zeta = \frac{3\pi^2k_0}{\sinh^2 \left(\pi k_0/2 \right)}
=  0.04636,
\ee
while the real part can only be obtained numerically:
\be
\Real \zeta = -0.8509. \label{Rek1}
\ee

Denoting $\epsilon A_0 \equiv a(t)$ the unscaled amplitude of 
the wobbling mode and recalling that $v = \epsilon V$, we express the amplitude equation
\eqref{freeamp} in terms of the original variables:
\be
\label{mainampeqfree}
i a_t = -\frac{\omega_0 \zeta}{2}   \, |a|^2a + \frac{\omega_0}{2} 
 v^2a +
\order{|a|^4}.
\ee
We are referring to Eq.\eqref{mainampeqfree} as
the ``master" amplitude equation. The master equation
contains solvability conditions at all orders 
[which arise simply by expanding 
the derivative $d/dt$ in powers of $\epsilon$]
but unlike any particular amplitude equation, it
is applicable {\it for all times}. 
 The master equation is
the final product of the asymptotic analysis; 
all the conclusions about the behaviour of the 
wobbler's amplitude shall be made on the basis 
of this equation.

The modulus of $a$ is governed by the equation
\begin{equation} \label{decay_law}
\frac{d}{d t}|a|^2 = -\omega_0 \Imag \zeta \, |a|^4 +
\order{|a|^5}.
\end{equation}
Since $\Imag \zeta>0$, the amplitude of
the wobbling is monotonically
decreasing with time: a constant emission of radiation damps the wobbler.
The decay law is straightforward from \eqref{decay_law}:
\begin{equation}
 \label{decay_rate}
|a(t)|^2 = \frac{|a(0)|^2} 
{1+ \omega_0 \Imag \zeta \, |a(0)|^2t }=
\frac{|a(0)|^2} 
{1+ 0.08030 \times |a(0)|^2t }.
\end{equation} 
When $a(0)$ is small, the decay becomes 
appreciable only after long times  $t \sim |a(0)|^{-2}$.  
The decay is slow; for times $t \gg 12.5 \times |a(0)|^{-2}$, 
Eq.\eqref{decay_rate} gives $|a| \sim t^{-1/2}$.

We have verified the above decay law in 
direct numerical simulations of the full partial differential 
equation \eqref{phi4}. As the
 initial conditions, we took $\phi(x,0)=\tanh x +2a_0 \sech x \tanh x$
with some real $a_0$ and
  $\phi_t(x,0)=0$. After a short initial transient,
  the solution was seen to settle to the curve  
  \eqref{decay_rate} with $|a(0)|$ close to $a_0$, see Fig.\ref{decay_figure}. 


\begin{figure}
  \includegraphics[height = 2.0in,width = 0.5\linewidth]{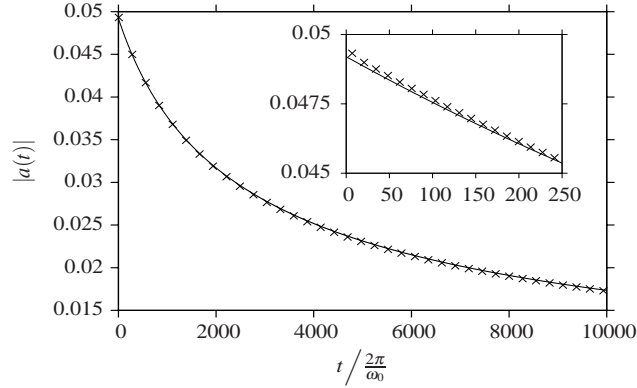}
  \caption{The decay of the free wobbling kink.  Crosses: 
    $|a(t)|$ as obtained from the direct numerical simulations
  of Eq.\eqref{phi4} with the initial conditions in the form
  $\phi=\tanh x +2a_0 \sech x \tanh x$,
  $\phi_t=0$, with $a_0=0.05$. Solid curve: equation 
  \eqref{decay_rate} with $|a(0)|=0.0492$.
  The inset shows the first 250 periods of oscillation;
  during this time the amplitude drops by less than $10\%$.}
  \label{decay_figure}
  \end{figure}

The  equation \eqref{mainampeqfree}  gives us the leading-order 
contributions 
to the frequency of the wobbling:
\begin{align}
\label{ampfreq}
\omega = \omega_0 \left[1-\tfrac{1}{2}v^2 + \tfrac{1}{2}
\Real \zeta \, |a|^2 + \order{|a|^4}\right],
\end{align}
with $\Real \zeta<0$ as in \eqref{Rek1}. 
 The $|a|^2$-term here is a 
nonlinear  shift from the linear frequency $\omega_0=\sqrt{3}$.
The $v^2$-term comes from the transverse  Doppler effect. 
We could have obtained this 
term simply by calculating the wobbling  frequency in the rest
frame and then multiplying the result by the relativistic time-dilation 
factor $\sqrt{1-v^2}$ (which becomes $1-\tfrac{1}{2}v^2$ for small
$v$).

\section{Conclusion}
\label{Conclusions}

One result of this project is the uniform
asymptotic expansion of the wobbling kink. In terms 
of the original variables, this expansion reads
\begin{multline} \label{phi_sum} 
\phi(x,t) = \tanh \left( \frac{1-3|a|^2}{\sqrt{1-v^2}} \xi \right) 
+ a \sech \xi \tanh \xi e^{i\omega_0 (t-v \xi)} + c.c. \\ 
+ 2 |a|^2\sech^2 \xi \tanh \xi  
+ a^2 f(\xi)e^{2i\omega_0 (t-v \xi)} + c.c. + \order{|a|^3}.
\end{multline}
Here $\xi=x-vt$ and $f(\xi)$ is given by
Eq.\eqref{f1}.
The first term describes the ``background", stationary, kink with the width
modified by the wobbling.
[Note that we have incorporated two $X_0 \sech^2 X_0$ terms 
of the sum \eqref{varphi2} into the  width of the kink.]
The second term is the wobbling mode itself; the third 
one accounts for further stationary deformation of the kink's shape,
and the last term represents the second-harmonic radiation
from the wobbler.

Our second result is the amplitude equation \eqref{mainampeqfree}
and the conclusion of the extreme longevity of the wobbling
mode which follows from \eqref{mainampeqfree}.
In view of its anomalously long lifetime,
the wobbling kink 
can be regarded as one of the fundamental nonlinear excitations of the 
$\phi^4$ theory, on par with the nonoscillatory kinks
and breathers.

\acknowledgments
O.O. was supported by funds provided by the NRF of South Africa
and the University of Cape Town. 
I.B. was supported by the NRF under
grants No. 65498 and 68536.

\end{document}